\documentclass[reprint,amssymb,amsmath,aip,cha,frontmatterverbose,floatfix]{revtex4-1}
\usepackage[T1]{fontenc}
\usepackage[latin9]{inputenc}
\usepackage{amsmath}
\usepackage{amssymb}
\usepackage{graphicx}
\usepackage[unicode=true,
 bookmarks=false,
 breaklinks=false,pdfborder={0 0 1},backref=section,colorlinks=false]
 {hyperref}
\usepackage{breakurl}

\makeatletter

\usepackage{dcolumn}
\usepackage{bm}
\usepackage{amsthm}
\usepackage{amsfonts}
\usepackage{float}
\usepackage{color}
 \usepackage{tcolorbox}
\usepackage{enumitem}
\setlist[itemize]{leftmargin=*}
\renewcommand{\textendash}{--}

\makeatother

\begin{document}

\title{JWvG:1749}

\title{Dynamics of a stochastic excitable system with slowly adapting feedback}

\author{Igor Franovi\'{c}, Serhiy Yanchuk, Sebastian Eydam, Iva Ba\v{c}i\'{c},
Matthias Wolfrum}
\begin{abstract}
We study an excitable active rotator with slowly adapting nonlinear
feedback and noise. Depending on the adaptation and the noise level,
this system may display noise-induced spiking, noise-perturbed oscillations,
or stochastic busting. We show how the system exhibits transitions
between these dynamical regimes, as well as how one can enhance or
suppress the coherence resonance, or effectively control the features
of the stochastic bursting. The setup can be considered as a paradigmatic
model for a neuron with a slow recovery variable or, more generally,
as an excitable system under the influence of a nonlinear control
mechanism. We employ a multiple timescale approach that combines the
classical adiabatic elimination with averaging of rapid oscillations
and stochastic averaging of noise-induced fluctuations by a corresponding
stationary Fokker-Planck equation. This allows us to perform a numerical
bifurcation analysis of a reduced slow system and to determine the
parameter regions associated with different types of dynamics. In
particular, we demonstrate the existence of a region of bistability,
where the noise-induced switching between a stationary and an oscillatory
regime gives rise to stochastic bursting.
\end{abstract}
\maketitle
\begin{quotation}
Recent years have witnessed a rapid expansion of stochastic models
for a wide variety of important physical and biological phenomena,
from sub-cellular processes and tissue dynamics, over large-scale
population dynamics and genetic switching to optical devices, Josephson
junctions, fluid mechanics and climatology. These studies have demonstrated
that the effects of noise manifest themselves on a broad range of
scales, but nevertheless display certain universal features. In particular,
the effects of noise may generically be cast into two groups. On the
one hand, the noise may enhance or suppress the features of deterministic
dynamics, while on the other hand, it may give rise to novel forms
of behavior, associated with the crossing of thresholds and separatrices,
or with stabilization of deterministically unstable states. The constructive
role of noise has been evinced in diverse applications, from neural
networks and chemical reactions to lasers and electronic circuits.
Classical examples of stochastic facilitation in neuronal systems
concern resonant phenomena, such as coherence resonance, where an
intermediate level of noise may trigger coherent oscillations in excitable
systems, as well as spontaneous switching between the coexisting metastable
states. In the present study, we show how the interaction of noise
and multiscale dynamics, induced by slowly adapting feedback, may
affect an excitable system. It gives rise to a new mode of behavior
based on switching dynamics, namely the stochastic bursting, and allows
for an efficient control of the properties of coherence resonance.
\end{quotation}

\section{Introduction}

Multiscale dynamics is ubiquitous in real-world systems. In neuron
models, for instance, the evolution of recovery or gating variables
is usually much slower than the changes of the membrane potential
\citep{Izhikevich2007,Gerstner2014}. At the level of neural networks,
certain mechanisms of synaptic adaptation, such as the spike timing-dependent
plasticity \citep{Abbott2005,Clopath2010,Popovych2013}, are slower
than the spiking dynamics of individual neurons. When modeling the
dynamics of semiconductor lasers \citep{Lang1980,Ludge2012,Soriano2013},
one similarly encounters at least two different timescales, one related
to the carriers' and the other to the photons' lifetime, whereby their
ratio can span several orders of magnitude. Investigating the dynamics
of such multiscale systems has lead to the development of a number
of useful asymptotic and geometric methods, see Refs.~\citep{Krupa1997,Lichtner2011,Desroches2012,Kuehn2015,Jardon-Kojakhmetov2019}
to name just a few.

Another ingredient inevitable in modeling real-world systems is noise,
which may describe the intrinsic randomness of the system, the fluctuations
in the embedding environment, or may derive from coarse-graining over
the degrees of freedom associated with small spatial or temporal scales
\citep{Haken1985,Lindner2004}. For instance, neuronal dynamics is
typically influenced by intrinsic sources of noise, such as the random
opening of ion channels, and by external sources, like the synaptic
noise \citep{Destexhe2012}. In chemical reactions, noise comprises
finite-size effects, while the stochasticity in laser dynamics reflects
primarily quantum fluctuations. In general, the impact of noise can
manifest itself by modification of the deterministic features of the
system, or by the emergence of qualitatively novel types of behavior,
induced by the crossing of thresholds or separatrices \citep{Forgoston2018}.

In the present paper, we study the effects of slowly adapting feedback
and noise on an excitable system. Excitability is a general nonlinear
phenomenon based on a threshold-like response of a system to a perturbation
\citep{Murray1989,Winfree2001,Lindner2004,Izhikevich2007}. An excitable
system features a stable \textquotedbl{}rest\textquotedbl{} state
intermitted by excitation events (firing), elicited by perturbations.
In the absence of a perturbation, such a system remains in the rest
state and a small perturbation induces a small-amplitude linear response.
If the perturbation is sufficiently strong, an excitable system reacts
by a large-amplitude nonlinear response, such as a spike of a neuron.
When an excitable system receives additional feedback or a stochastic
input, or is coupled to other such systems, new effects may appear
due to the self- or noise-induced excitations, as well as excitations
from the neighboring systems. Such mechanisms can give rise to different
forms of oscillations, patterns, propagating waves, and other phenomena
\citep{Lindner2004,Pikovsky1997,VallesCodina2011,Ermentrout2001,LueckenRosinWorlitzerEtAl2017,Franovic2018,Bacic2018,FranovicTodorovic2015,FranovicPerc2015,Yanchuk2019}.

Our focus is on a stochastic excitable system subjected to a slow
control via a low-pass filtered feedback 
\begin{align}
\dot{v} & =f(v,\mu)+\sqrt{D}\xi(t),\label{eq:v}\\
\dot{\mu} & =\varepsilon(-\mu+\eta g(v)),\label{eq:mu}
\end{align}
where $\varepsilon\gtrsim0$ is a small parameter that determines
the timescale separation between the fast variable $v(t)$ and the
slow feedback variable $\mu(t)$. The fast dynamics $\dot{v}=f(v,0)$
is excitable and is influenced by the Gaussian white noise $\xi(t)$
of variance $D$. Moreover, the slow feedback variable $\mu$ controls
its excitability properties. The parameter $\eta$ is the control
gain, such that for $\eta=0$ one recovers a classical noise-driven
excitable system \citep{Lindner2004}. An important example of a system
conforming to \eqref{eq:v}\textendash \eqref{eq:mu} for $\eta\neq0$
is the Izhikevich neuron model \citep{Izhikevich2004}, where the
stochastic input to the fast variable would describe the action of
synaptic noise.

Here we analyze a simple paradigmatic example from the class of systems
\eqref{eq:v}\textendash \eqref{eq:mu}, where the excitable local
dynamics is represented by an active rotator 
\[
\dot{\varphi}=I_{0}-\sin\varphi\quad\textrm{with}\quad\varphi\in S^{1}.
\]
The latter undergoes a saddle-node infinite period (SNIPER) bifurcation
at $|I_{0}|=1$, turning from excitable ($|I_{0}|\lesssim1$) to oscillatory
regime $|I_{0}|>1$, see \citep{Strogatz1994}. The adaptation is
represented by a positive periodic function $g(\varphi)=1-\sin\varphi$,
such that the complete model reads 
\begin{align}
\dot{\varphi} & =I_{0}-\sin\varphi+\mu+\sqrt{D}\xi(t),\label{eq1a}\\
\dot{\mu} & =\varepsilon\left(-\mu+\eta\left(1-\sin\varphi\right)\right).\label{eq1b}
\end{align}
In the presence of feedback, the noiseless dynamics of the active
rotator depends not only on $I_{0}$, but is affected by the term
$I_{0}+\mu$ involving the control variable $\mu(t)$, which can induce
switching between the excitable equilibrium ($|I_{0}+\mu|<1$) and
the oscillatory regime ($|I_{0}+\mu|>1$). This adaptation rule provides
a positive feedback for the spikes and oscillations, since $\mu$
increases when $\varphi(t)$ is oscillating and drives the system
towards the oscillatory regime, while in the vicinity of the equilibrium
($\sin\varphi\approx1$) the control signal effectively vanishes.

We examine how the behavior of \eqref{eq1a}-\eqref{eq1b} is influenced
by the noise level $D$ and the control gain $\eta$, determining
the phase diagram of dynamical regimes in terms of these two parameters.
The first part of our results in Sec.~\ref{sec:deterministic} concerns
the noise-free system $D=0$, where we employ a combination of two
multiscale methods, namely adiabatic elimination in the regime where
the fast subsystem has a stable equilibrium and the averaging approach
when the fast subsystem is oscillatory. As a result, we obtain a reduced
slow system that is capable of describing both the slowly changing
fast oscillations and the slowly drifting equilibrium, as well as
the transitions between these regimes. The bifurcation analysis of
this slow system reveals the emergence of a bistability between the
fast oscillations and the equilibrium for sufficiently large $\eta$.

The second part of our results, presented in Sec.~\ref{sec:3}, addresses
the multiscale analysis of the dynamics in the presence of noise ($D\neq0$).
Instead of deterministic averaging, we apply the method of \emph{stochastic
averaging} \citep{Shilnikov2008,Pavliotis2008,Galtier2012,Lucken2016,Bacic2018},
where the distribution density for the fast variable obtained from
a stationary Fokker-Plank equation is used to determine the dynamics
of the slow flow. In this way, we obtain a deterministic slow dynamics
for which one can perform a complete numerical bifurcation analysis
with respect to $D$ and $\eta$. In section \ref{sec:4} we investigate
the effects of stochastic fluctuations on the slow dynamics, which
vanish in the limit of infinite timescale separation $\varepsilon\rightarrow0$
employed in Sec.~\ref{sec:3}. The effect of a slowly adapting feedback
on the coherence resonance is shown by extracting from numerical simulations
the coefficient of variation of the spike time distribution in the
excitable regime. In particular, we compare the results for small
positive $\varepsilon$ with the case of infinite time scale separation,
where we use the stationary but noise dependent $\mu$ obtained in
the preceeding section. The noise-induced switching dynamics in the
bistability region is demonstrated by numerical simulations showing
an Eyring-Kramers type of behavior.

\begin{figure}
\centering{}\includegraphics[width=0.8\columnwidth]{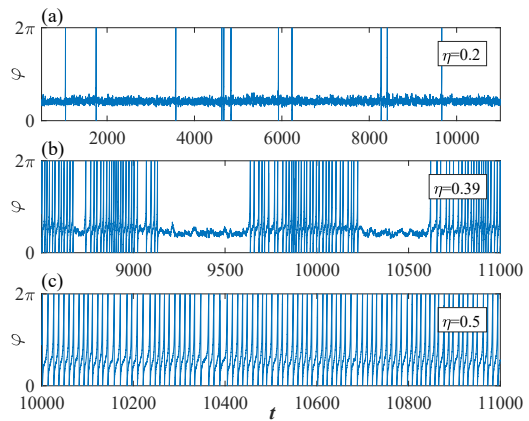}\caption{\label{fig:1} Different dynamical regimes in the stochastic excitable
system subjected to a slow control via a low-pass filtered feedback
\eqref{eq1a}-\eqref{eq1b} with $\varepsilon=0.005,\,D=0.008,$ and
different choices of the control gain $\eta$: noise-induced spiking
(a), stochastic bursting (b), and noise-perturbed spiking (c).}
\end{figure}

In terms of the different dynamical regimes, our study of stochastic
dynamics reveals three characteristic $(D,\eta)$ regions featuring
noise-induced spiking, noise-perturbed spiking and stochastic busting,
see Figure~\ref{fig:1}. We show that by varying the control gain
within the region of noise-induced spiking, one can enhance or suppress
the coherence resonance, while within the bistability region, one
can efficiently control the properties of stochastic bursting. The
following sections provide a detailed analysis of the described phenomena.

\section{Slow-fast analysis of the deterministic dynamics}

\label{sec:deterministic}

In this Section, we analyze the system \eqref{eq1a}\textendash \eqref{eq1b}
in the absence of noise $(D=0)$ 
\begin{align}
\dot{\varphi(t)} & =I_{0}-\sin\varphi(t)+\mu(t),\label{eq1a-D}\\
\dot{\mu(t)} & =\varepsilon\left(-\mu(t)+\eta\left(1-\sin\varphi(t)\right)\right),\label{eq1b-D}
\end{align}
considering the limit $\varepsilon\rightarrow0$ within the framework
of singular perturbation theory. The dynamics on the fast timescale
is described by the so-called layer equation, obtained from \eqref{eq1a-D}\textendash \eqref{eq1b-D}
by setting $\varepsilon=0$ 
\begin{equation}
\dot{\varphi}(t)=I_{0}+\mu-\sin\varphi(t),\label{eq:lay}
\end{equation}
whereby $\mu$ acts as a parameter.

\subsection{Dynamics for $\mu<1-I_{0}$: adiabatic elimination}

\begin{figure}
\begin{centering}
\includegraphics[width=0.8\columnwidth]{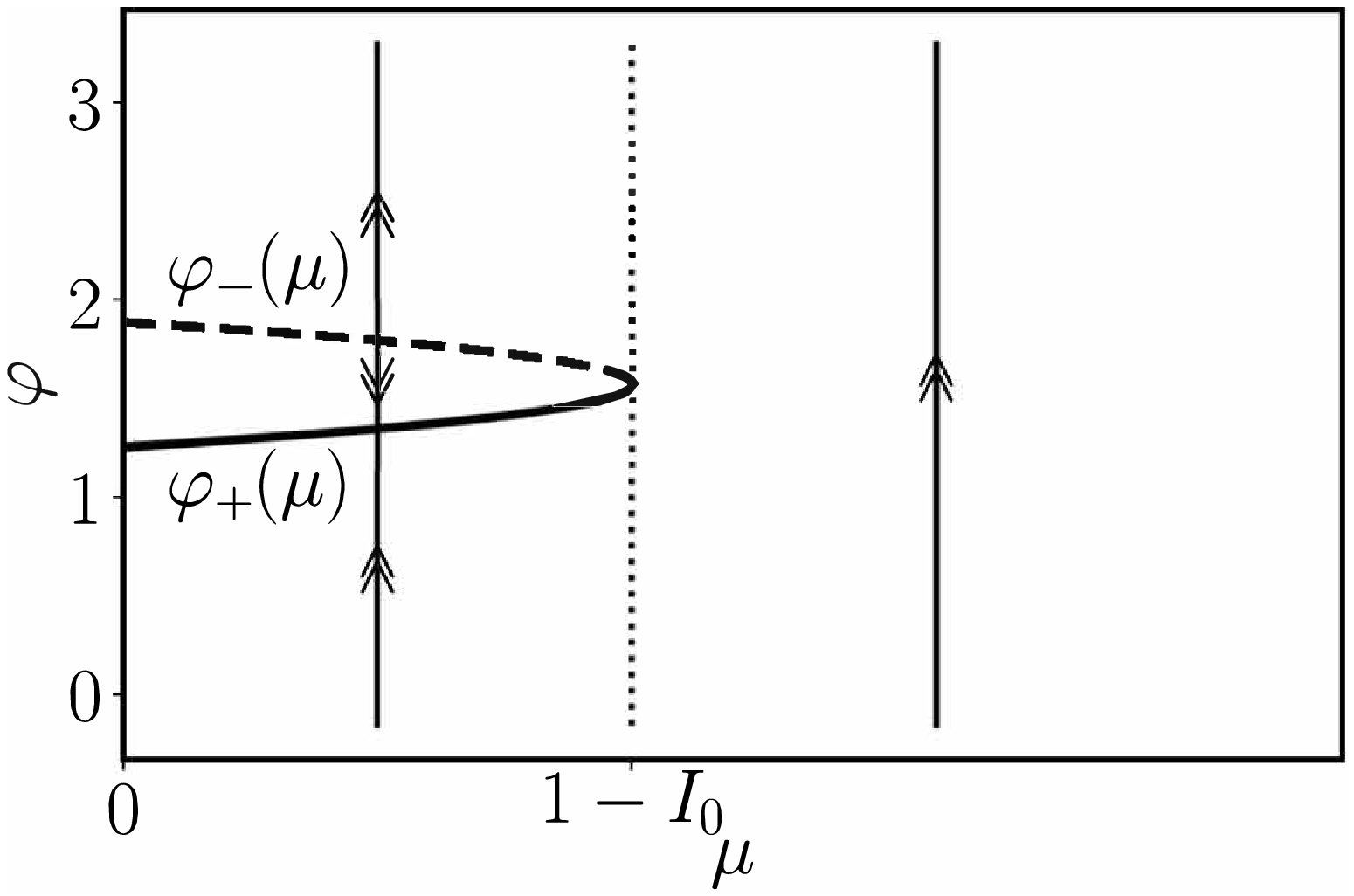}
\par\end{centering}
\caption{\label{fig:slow-fast-simple} Critical manifold and fast dynamics
of system \eqref{eq1a-D}\textendash \eqref{eq1b-D}. For $\mu<1-I_{0}$
the fast dynamics converges to the stable branch of the critical manifold,
while for $\mu>1-I_{0}$, it is oscillatory with periodic rotation
of the phase $\varphi$.}
\end{figure}

In the case $\mu<1-I_{0}$, the layer equation (\ref{eq:lay}) possesses
two equilibria
\begin{equation}
\varphi_{+}(\mu)=\arcsin(I_{0}+\mu),\,\,\,\varphi_{-}(\mu)=\pi-\varphi_{+}(\mu),\label{eq:stbra}
\end{equation}
where $\varphi_{+}$ is stable and $\varphi_{-}$ is unstable. Considering
them as functions of the parameter $\mu$, the equilibria give rise
to two branches, which merge in a fold at $\mu=1-I_{0}$, see Fig.~\ref{fig:slow-fast-simple}.
Equivalently, the set of equililbria of the fast subsystem 
\begin{equation}
\left\{ (\varphi,\mu):\,\sin\varphi=I_{0}+\mu\right\} \label{eq:critman}
\end{equation}
comprises the critical manifold of \eqref{eq1a-D}\textendash \eqref{eq1b-D},
with the stable part $\varphi_{+}(\mu)$ and the unstable part $\varphi_{-}(\mu)$..

Hence, for $\mu<1-I_{0}$ the trajectories are rapidly attracted towards
the stable branch of the critical manifold, along which for positive
$\varepsilon$ they slowly drift. In order to describe this slow dynamics,
we rescale time $T=\varepsilon t$ and obtain 
\begin{align}
\varepsilon\varphi^{\prime}(T) & =I_{0}+\mu(T)-\sin\varphi(T),\label{eq3a}\\
\mu^{\prime} & (T)=-\mu(T)+\eta(1-\sin\varphi(T)),\label{eq3b}
\end{align}
where the prime denotes the derivative with respect to the slow time
$T$. Setting $\varepsilon=0$, we can directly eliminate the term
$\sin\varphi=I_{0}+\mu$ and obtain the equation for the slow dynamics
on the critical manifold
\begin{equation}
\mu^{\prime}(T)=-\mu(T)+\eta(1-I_{0}-\mu(T)).\label{eq:slow1}
\end{equation}

\subsection{Dynamics for $\mu>1-I_{0}$: averaging fast oscillations}

For $\mu>1-I_{0}$, there is no stable equilibrium of the fast subsystem
\eqref{eq:lay}, see Fig.~\ref{fig:slow-fast-simple}. Instead, one
finds periodic oscillations 
\begin{align}
\varphi_{\mu}(t)=2\arctan\frac{1+\Omega(\mu)\tan\frac{t}{2}\Omega(\mu)}{I_{0}+\mu}\label{eq:po}
\end{align}
with the $\mu$-dependent frequency 
\[
\Omega(\mu)=\sqrt{(I_{0}+\mu)^{2}-1}.
\]
In this case, the fast oscillations $\varphi_{\mu}(t)$ should be
averaged in order to obtain the dynamics of the slow variable $\mu(T)$.
A rigorous formal derivation is provided in Appendix~A, finally arriving
at
\begin{align}
\mu^{\prime} & (T)=-\mu(T)+\eta(1-I_{0}-\mu(T)+\Omega(\mu(T))).\label{eq:slow2}
\end{align}
Here we give a simplified explanation of the averaging procedure.
First, we substitute the fast-oscillating solution $\varphi=\varphi_{\mu}(t)$
of the layer equation into the equation for the slow variable (\ref{eq3b}):
\[
\mu^{\prime}(T)=-\mu(T)+\eta(1-\sin\varphi_{\mu}(t)).
\]
Since the term $\sin(\cdot)$ is fast oscillating, the last equation
can be averaged over the fast timescale $t,$which leads to
\begin{equation}
\mu^{\prime}(T)=-\mu(T)+\eta\left(1-\left\langle \sin\varphi_{\mu}(t)\right\rangle _{t}\right).\label{eq:ttt}
\end{equation}
The average $\left\langle \sin\varphi_{\mu}(t)\right\rangle _{t}$
can be found by integrating \eqref{eq:lay} over the period 
\begin{equation}
\langle\dot{\varphi}(t)\rangle_{t}=\Omega(\mu)=I_{0}+\mu-\langle\sin\varphi_{\mu}(t)\rangle_{t}.\label{eq:pav}
\end{equation}
Hence, by substituting 
\[
\langle\sin\varphi_{\mu}(t)\rangle_{t}=I_{0}+\mu(T)-\Omega(\mu(T))
\]
into (\ref{eq:ttt}) we obtain the slow averaged dynamics (\ref{eq:slow2}).
\begin{figure}[t]
\centering \includegraphics[width=0.99\linewidth]{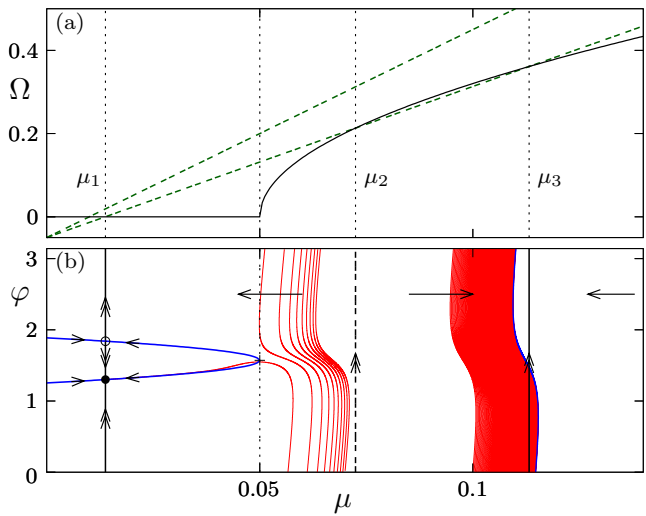}
\caption{\label{Fig:SlowFast-1} (a) Graphical solution of the fixed point
equation \eqref{eq:fp}: $\Omega(\mu)$ according to \eqref{eq15}
(black) and the righ-hand side of \eqref{eq:fp} for different choices
of $\eta$. One finds from one to three fixed points depending on
$\eta$. (b) Scheme of the slow-fast dynamics of system \eqref{eq1a-D},\eqref{eq1b-D}
with parameters $I_{0}=0.95$ and $\eta=0.38$ and the numerical sample
trajectories for $\varepsilon=0.005$ (red). For $\mu<1-I_{0}$, trajectories
are attracted to the stable branch of the slow manifold (blue curve)
and subsequently slowly drift toward the stable fixed point $(\varphi_{+}(\mu_{1}),\mu_{1})$
(black dot). For $\mu>1-I_{0}$, the sample trajectories show fast
oscillations in $\varphi$ with a slow average drift in $\mu$ in
the direction indicated by the arrows.}
\end{figure}

\subsection{Combined dynamics of the slow variable}

Summarizing the results so far, the equation (\ref{eq:slow1}) describes
the dynamics of the slow variable for $\mu<1-I_{0}$, while the equation
(\ref{eq:slow2}) holds for $\mu>1-I_{0}$. These two equations can
be conveniently combined into a single equation of the form (\ref{eq:slow2})
by extending the definition of the frequency $\Omega(\mu)$ as follows
\begin{equation}
\Omega(\mu)=\left\{ \begin{array}{lr}
0, & \mu<1-I_{0}\\
\sqrt{(I_{0}+\mu)^{2}-1}, & \mu>1-I_{0}
\end{array}\right..\label{eq15}
\end{equation}
Hence, the slow dynamics is described by the scalar ordinary differential
equation on the real line (\ref{eq:slow2}), and, as a result, the
only possible attractors are fixed points, which are given by the
zeros of the right-hand side: 
\begin{align}
\Omega(\mu)=\frac{\eta+1}{\eta}\mu+I_{0}-1\label{eq:fp}
\end{align}
Geometrically, they are points of intersection of the frequency profile
$\Omega(\mu)$ with the line $\frac{\eta+1}{\eta}\mu+I_{0}-1$, see
Fig.~\ref{Fig:SlowFast-1}(a). In particular, one can check that
there is always one fixed point 
\begin{equation}
\mu_{1}=\frac{\eta(1-I_{0})}{1+\eta}<1-I_{0}
\end{equation}
for which $\Omega(\mu_{1})=0$, such that it corresponds to a pair
of equilibria on the critical manifold \eqref{eq:critman}. Since
$\mu_{1}$ is stable for the slow dynamics, the point $(\varphi_{+}(\mu_{1}),\mu_{1})$
is also a stable equilibrium for the original system \eqref{eq1a-D}\textendash \eqref{eq1b-D}
with small $\varepsilon$. The other two fixed points of the slow
equation 
\begin{equation}
\mu_{2,3}=\frac{\eta\left(1+\eta-I_{0}\mp\sqrt{(\eta+I_{0})^{2}-1-2\eta}\right)}{1+2\eta}
\end{equation}
with $\Omega(\mu_{2,3})>0$ appear in a saddle-node bifurcation at
\begin{equation}
\eta_{\textrm{sn}}=1-I_{0}+\sqrt{2(1-I_{0})},\label{eq:eta_c}
\end{equation}
and correspond to a pair of periodic orbits of the layer equation
\eqref{eq:lay}.


In Fig. \ref{Fig:SlowFast-1}(b) we show schematically the results
of our slow-fast analysis for $I_{0}=0.95$ and $\eta=0.38$. For
the chosen parameter values there are two stable regimes: the fixed
point $(\varphi_{+}(\mu_{1}),\mu_{1})$ and a fast oscillation with
$\langle\mu(t)\rangle_{t}\approx\mu_{3}$.

\begin{figure}[t]
\centering \includegraphics[width=0.99\linewidth]{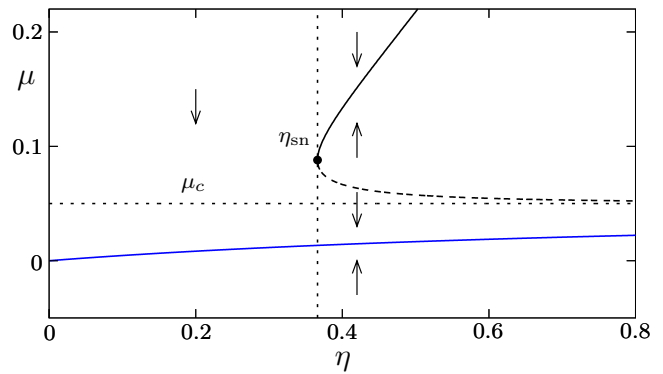} \caption{\label{Fig:muD0} Fixed points of the slow dynamics \eqref{eq:slow2}
for varying control gain $\eta$. The values $\mu_{2,3}$ on the upper
branch (black curve) correspond to periodic orbits of the layer equation
\eqref{eq:lay}, while $\mu_{1}$ (blue curve) is the branch of fixed
points; solid and dashed lines indicate stable and unstable solutions,
respectively. The direction of the motion in $\mu(T)$ is indicated
by the arrows. The dotted lines indicate the onset of bistability
for $\eta=\eta_{\textrm{sn}}$ and the transition at $\mu_{c}=1-I_{0}$
from equilibria to periodic orbits.}
\end{figure}

Finally, Fig.~\eqref{Fig:muD0} presents the bifurcation diagram
of the fixed points of the slow dynamics with respect to the control
gain $\eta$. One observes that there is always one branch of stable
fixed points corresponding to the steady state, and two stable fixed
points corresponding to fast oscillations for $\eta>\eta_{\textrm{sn}}$.

\section{Slow-fast analysis of the dynamics with noise}

\label{sec:3}

\begin{figure}[t]
\centering \includegraphics[width=0.99\linewidth]{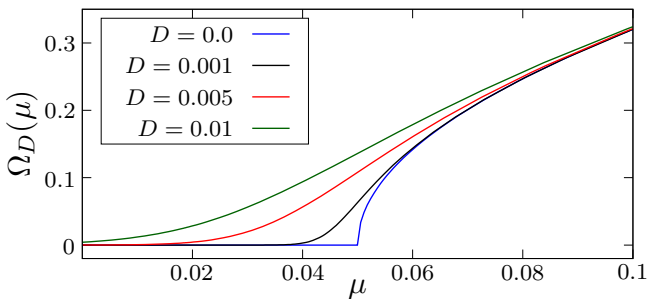} \caption{\label{Fig:OmegaD-1} Average frequency of the fast dynamics \eqref{eq1a}
given by \eqref{eq20}- \eqref{eq21} using numerical solutions of
the stationary Fokker-Planck equation \eqref{eq:FoPla}, where $\mu$
acts as a time independent parameter and fixed $I_{0}=0.95$.}
\end{figure}
In this section, we consider the dynamics of system \eqref{eq1a}\textendash \eqref{eq1b}
in the presence of noise ($D>0$). In analogy to the noise-free case,
one can use the limit $\varepsilon\rightarrow0$ and employ the \textit{stochastic
average} 
\[
\langle\sin\varphi(t)\rangle_{t}=\lim_{t\longrightarrow\infty}\frac{1}{t}\int_{0}^{t}\sin\varphi(t)\mathrm{d}t'
\]
for solutions of the stochastic fast equation 
\begin{equation}
\dot{\varphi}(t)=I_{0}+\mu-\sin\varphi(t)+\sqrt{D}\xi(t)\label{eq:layN}
\end{equation}
to approximate the slow dynamics in \eqref{eq3b} by 
\begin{align}
\mu^{\prime}(T) & =-\mu(T)+\eta(1-\langle\sin\varphi(t)\rangle_{t}).\label{eq:23}
\end{align}
To this end, we consider the \emph{stationary probability density
distribution} $\rho(\varphi;\mu,D)$ for the fast noisy dynamics \eqref{eq1a},
which for fixed control $\mu$ and noise intensity $D$ is given as
a solution to the stationary Fokker-Planck equation 
\begin{equation}
\frac{D}{2}\partial_{\varphi\varphi}\rho-\partial_{\varphi}\left[(I_{0}+\mu-\sin\varphi)\rho\right]=0,\label{eq:FoPla}
\end{equation}
together with the periodic boundary conditions $\rho(0)=\rho(2\pi)$
and the normalization 
\begin{equation}
\int_{0}^{2\pi}\rho(\varphi;\mu,D)\mathrm{d}\varphi=1.\label{eq:BC}
\end{equation}
From this we can calculate the average 
\begin{align}
\langle\sin\varphi(t)\rangle_{t} & =\int_{0}^{2\pi}\rho(\varphi;\mu,D)\sin\varphi\mathrm{d}\varphi\label{eq20}
\end{align}
and obtain the mean frequency 
\begin{align}
\Omega_{D}(\mu) & =I_{0}+\mu-\langle\sin\varphi(t)\rangle_{t},\label{eq21}
\end{align}
which depends via \eqref{eq20} both on $D$ and $\mu$. Taking into
account (\ref{eq:23}) and (\ref{eq21}), the equation for the slow
dynamics of $\mu(T)$ reads
\begin{equation}
\mu^{\prime}(T)=-\mu(T)+\eta(1-I_{0}-\mu+\Omega_{D}(\mu(T))),\label{eq:slow3}
\end{equation}
i.e. it is of the same form as in the deterministic case (\ref{eq:slow2}).
The corresponding fixed point equation for the stationary values of
$\mu$ with respect to the slow dynamics is given by \eqref{eq:fp}.

The stationary Fokker-Planck equation \eqref{eq:FoPla} can be solved
directly by integral expressions, see Appendix B\textbf{.} In particular,
for $D=0$ we readily recover the results for periodic averaging from
the previous section. However, for small non-vanishing $D$, the integrals
become difficult to evaluate numerically and we preferred to solve
\eqref{eq:FoPla} as a first-order ODE boundary value problem with
the software AUTO \citep{Doedel2006}, which provides numerical solutions
to boundary value problems by collocation methods together with continuation
tools for numerical bifurcation analysis.

\begin{figure}[t]
\centering \includegraphics[width=0.99\linewidth]{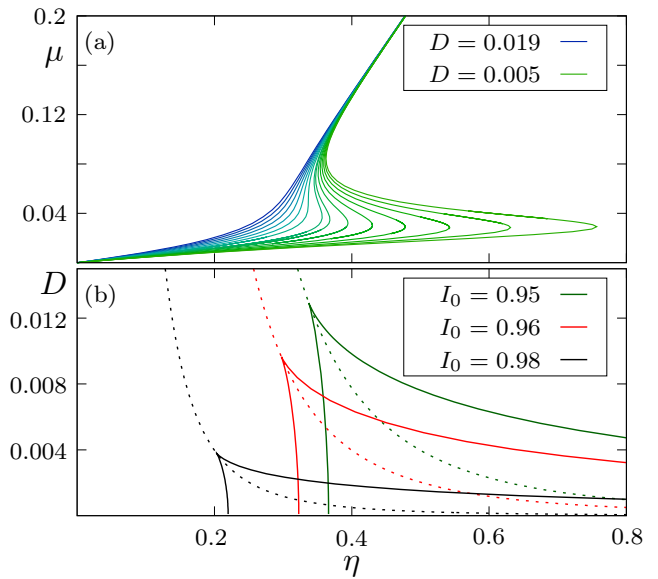}
\caption{(a) Branches of fixed points $\mu^{\ast}(\eta)$ of the slow dynamics
(\ref{eq:slow3}) for a set of noise values $D=0.005,0.006,\dots,0.019$,
and $I_{0}=0.95$, calculated from \eqref{eq:fp} together with the
stationary Fokker-Planck equation \eqref{eq:FoPla}. (b) Two-dimensional
bifurcation diagrams in terms of $\eta$ and $D$ for three different
values of $I_{0}$ show the curves of fold bifurcations, which meet
at the cusp point. Dashed curves indicate the case where $\mu=\mu_{c}=1-I_{0}$.}
\label{Fig:Cusp}
\end{figure}

\begin{figure}[t]
\centering \includegraphics[width=0.99\linewidth]{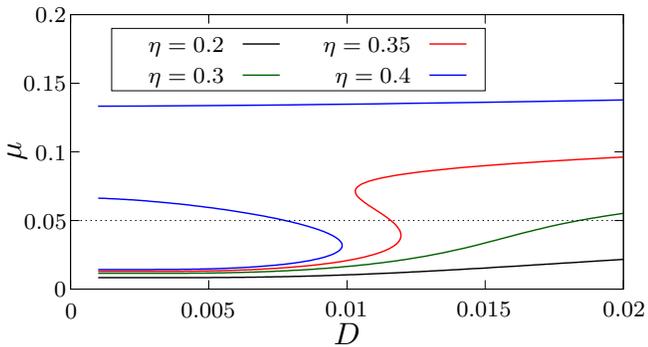} \caption{Branches of fixed points $\mu^{\ast}(D)$ of the slow dynamics (\ref{eq:slow3})
for a set of control gain values $\eta\in\left\{ 0.2,0.3,0.35,0.4\right\} $
and fixed $I_{0}=0.95$, calculated from \eqref{eq:fp} together with
the stationary Fokker-Planck equation \eqref{eq:FoPla}.}
\label{Fig:muD}
\end{figure}

In Fig.~\ref{Fig:OmegaD-1} are shown the numerically obtained effective
frequencies $\Omega_{D}(\mu)$ for different noise levels $D$. Solving
the stationary Fokker-Planck equation \eqref{eq:FoPla} together with
the fixed point equation for $\mu(T)$ \eqref{eq:fp}, we obtain for
fixed values of $D$ and varying control gain $\eta$ branches of
stationary solutions $(\mu^{\ast},\rho(\varphi;\mu^{\ast},D))$, see
Fig.~\ref{Fig:Cusp}(a). For small noise intensities, these branches
are folded, which indicates the coexistence of up to three stationary
solutions, similar as in the noise-free case. Alternatively, we can
also fix $\eta$ and obtain branches for varying $D$, see Fig.~\ref{Fig:muD}.
For small $\eta$ they are monotonically increasing, while for larger
$\eta$ they are folded. For $\eta_{\textrm{sn}}<\eta$ there are
two separate branches, emanating from the three solutions of \eqref{eq:fp}
at $D=0$.

Continuation of the folds in the $(\eta,D)$ parameter plane provides
the curves outlining the boundaries of the bistability region. Fig.~\ref{Fig:Cusp}(b)
shows that the two branches of folds meet at the cusp point $(\eta_{\textrm{cu}},D_{\textrm{cu}})$.
One of the branches approaches for $D\rightarrow0$ the value $\eta=\eta_{\textrm{sn}}$,
which we have calculated in \eqref{eq:eta_c}, while the other one
diverges to infinite values of $\eta$. When $I_{0}$ approaches the
critical value $I_{0}=1,$ the cusp point shifts to a smaller noise
intensity $D$, such that the region of bistability decreases.

Note that for $D>0$ all the average frequencies satisfy $\Omega_{D}>0$
such that a clear distinction between the stationary and the oscillatory
regime of the fast dynamics is no longer possible. However, one can
compare the critical value of the deterministic fast dynamics 
\begin{equation}
\mu_{c}=1-I_{0}\label{eq:muc}
\end{equation}
with the corresponding stationary value $\mu^{*}$ of the slow variable
from (\ref{eq:slow3}) to distinguish between a regime of noise-induced
oscillations and oscillations derived from the deterministic part
of the dynamics. If $\mu^{*}<\mu_{c}$, the oscillations are noise-induced
and have the form of rare spikes, see Fig.~\ref{fig:1}(a),while
for $\mu^{*}>\mu_{c}$ the deterministic oscillations are prevalent,
see Fig.~\ref{fig:1}(c).

It turns out that the curves where the stationary values of $\mu$
satisfy the condition $\mu=\mu_{c}$, shown dashed in Fig.~\ref{Fig:Cusp}(b),
pass exactly through the corresponding cusp point and inside the bistability
region refer to the unstable solutions given by the middle part of
the S-shaped curves in Fig. \ref{Fig:Cusp}(a). From this we conclude
that changing the parameters across this line outside the bistability
region results in a gradual transition between the regimes of noise-induced
oscillations and the deterministic-driven oscillations, while a hysteretic
transition between the two stable regimes is obtained at the boundary
of the bistability region. Moreover, for finite timescale separation
$\varepsilon>0,$ there can be transitions between the two stable
regimes also within the bistability region, which are induced by the
stochastic fluctuations. In the following section we study in detail
how the region of bistability found for the singular limit $\varepsilon\rightarrow0$
also affects the dynamics of the original system in case of a finite
timescale separation.

\section{\label{sec:4}Effects of fluctuations and finite timescale separation}

\begin{figure}[t]
\centering \includegraphics[width=0.99\linewidth]{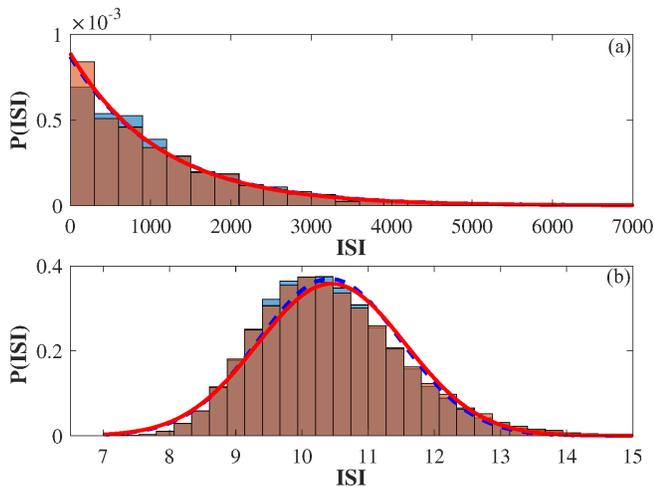} \caption{Histograms of inter spike intervals of the phase variable for control
gain $\eta=0.2$ (top panel) and $\eta=0.5$ (bottom panel), obtained
from numerical simulations of the full system \eqref{eq1a}\textendash \eqref{eq1b}
with $\varepsilon=0.005$ (orange) and in the limit of infinite timescale
separation (blue), using \eqref{eq:layN} with the stationary $\mu(T)\equiv\mu_{D}$
determined from the stationary Fokker-Planck equation \eqref{eq:FoPla}.
Solid red and dashed blue curves represent fits to an exponential
decay (a) and a Gaussian (b) for the histograms concerning the full
system and the limit of infinite scale separation, respectively.}
\label{fig9}
\end{figure}

The two basic deterministic regimes of the fast dynamics, which are
the excitable equilibrium and the oscillations, induce in a natural
way the two corresponding states of the system with noise and small
$\varepsilon>0$, namely
\begin{itemize}
\item Noise-induced spiking, characterized by a Poissonian-like distribution
of inter-spike intervals (ISIs), see Fig.~\ref{fig9}(a);
\item Noisy oscillations, involving a Gaussian-like distribution of the
ISIs, centered around the deterministic oscillation period, see Fig.~\ref{fig9}(b).
\end{itemize}
These states are found for sufficiently small or large values of $\eta$,
respectively, where only a corresponding single branch of the deterministic
system is available and the fluctuations of $\mu$ around its average
value have no substantial impact on the dynamics, cf. the blue and
orange distributions in Fig.~\ref{fig9}. For sufficiently large
noise levels above the cusp $(D>D_{cu})$ and intermediate values
of $\eta,$ one observes a gradual transition between these two regimes.
However, for smaller noise $D<D_{cu}$, allowing for the existence
of the region of bistability (cf. Fig.~\ref{Fig:Cusp}(b)), new regimes
of stochastic dynamics can emerge, namely:
\begin{itemize}
\item Enhanced coherence resonance, where a noise-induced dynamical shift
of the excitability parameter $I_{0}+\mu_{D}$ is self-adjusted close
to criticality;
\item Noise-induced switching between the two coexisting regimes in the
bistability region, see Fig.~\ref{fig:1}(b).
\end{itemize}

\subsection{Enhanced coherence resonance}

\begin{figure}[t]
\centering \includegraphics[width=0.99\linewidth]{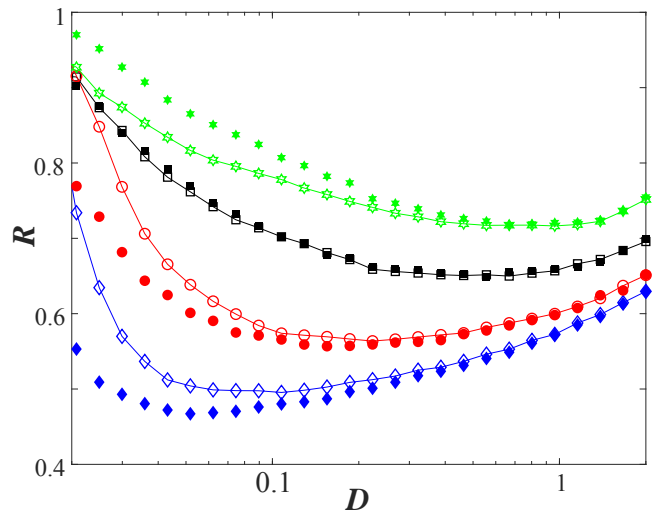} \caption{Enhancement or suppression of coherence resonance by a slowly adapting
feedback control. The connected lines with empty symbols refer to
$R(D)$ dependencies for the full system \ref{eq1a}-\ref{eq1b} at
different values of the control gain: $\eta=-0.2$ (green hexagonals),
$\eta=0$ (black squares), $\eta=0.2$ (red circles), and $\eta=0.3$
(blue diamonds), having fixed $I_{0}=0.95,\varepsilon=0.005$. The
unconnected filled symbols indicate the corresponding $R(D)$ dependencies
obtained from numerical simulations of the layer equation\ref{eq:layN}
with stationary $\mu^{\ast}(D)$.}
\label{fig11}
\end{figure}

The phenomenon of coherence resonance \citep{Pikovsky1997,Lindner1999,Makarov2001},
where the regularity of noise-induced oscillations becomes maximal
at an intermediate noise level, is well-known for noisy excitable
systems such as the fast equation \eqref{eq:layN} without adaptation,
i.e. for $\eta=0$ and therefore also $\mu=0$. For values of the
control gain $0<\eta<\eta_{\textrm{cu}}$ below the region of bistability,
the control leads to a substantially enhanced coherence resonance.
This effect can be quantified by studying the noise dependence of
the coefficient of variation of the inter spike intervals. For a given
noisy trajectory of \eqref{eq:layN},the spiking times $t_{k}$ are
defined as the first passage times $\varphi(t_{k})=2\pi k$, $k\in\mathbb{N}$
with corresponding inter spike intervals $\tau_{k}=t_{k}-t_{k-1}$.
The coefficient of variation of their distribution is defined as 
\begin{align}
R(D) & =\frac{\sqrt{\langle\tau_{k}^{2}\rangle-\langle\tau_{k}\rangle^{2}}}{\langle\tau_{k}\rangle}.\label{eq30}
\end{align}
For \eqref{eq:layN} with a fixed $\mu$, the latter can be determined
from direct numerical simulations. However, inserting for $\mu$ the
corresponding stochastic averages $\mu^{*}(D;\eta)$ obtained in Section
shows a strongly nonlinear dependence both on $\eta$ and $D$, see
also Figs. \ref{Fig:Cusp}(a) and \ref{Fig:muD}. In particular, the
strongly nonlinear dependence on $D$ for $\eta$ slightly below the
cusp value $\eta_{\textrm{cu}}$ has a substantial impact on the resonant
behavior reflected in the form of $R(D)$. In Fig.~\ref{fig11},
we show the $R(D)$ dependence for different values of the control
gain $\eta$, comparing the numerical results for the fast subsystem
\eqref{eq:layN} with inserted stationary values $\mu^{*}(D;\eta)$,
to numerical simulations of \eqref{eq1a}-\eqref{eq1b} for $\varepsilon=0.005$.
While for $0<\eta<\eta_{cu}$ one finds that the coherence resonance
can be substantially enhanced, cf. for example the $R(D)$ dependencies
for $\eta=0$ and $\eta=0.3$, note that by introducing the negative
values of the control gain $\eta,$ the resonant effect can be readily
suppressed. This implies that the adaptive feedback we employ provides
an efficient {\em control of coherence resonance}. Such an effect
has already been demonstrated in \citep{Aust2010,Kouvaris2010,Janson2004}by
using a delayed feedback control of Pyragas type. However, this control
method requires the feedback delay time as an additional control parameter
to be well adapted to the maximum resonance frequency..

\subsection{Bursting behavior due to noise-induced switching}

\begin{figure}[t]
\centering \includegraphics[width=0.99\linewidth]{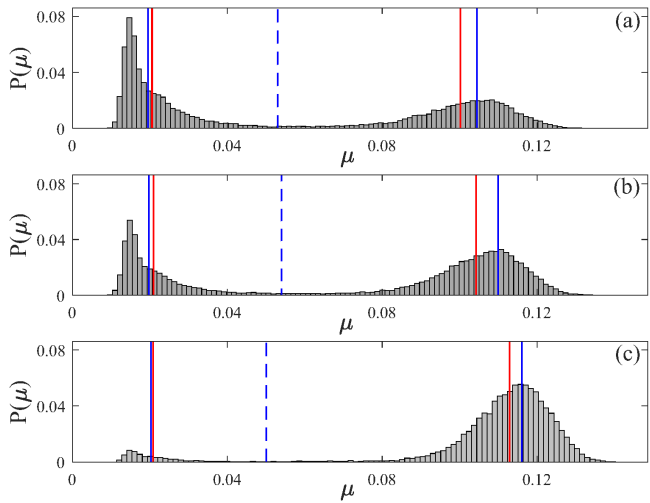} \caption{Stationary distributions $P(\mu)$, sampled from numerical simulations
of \eqref{eq1a}\textendash \eqref{eq1b} with $\varepsilon=0.005$.
Parameters $\eta=0.37$ in (a), $\eta=0.373$ in (b) and $\eta=0.38$
in (c) and fixed noise level $D=0.009$ lie inside the bistability
region from Fig.~\ref{Fig:Cusp}(b). Blue vertical lines indicate
the fixed points of $\mu$ from the stationary Fokker-Planck equation
\eqref{eq:FoPla} together with the fixed point equation \eqref{eq:fp}
of the slow dynamics. Red vertical lines indicate the mean values
of all $\mu$ in $P(\mu)$ below and of all $\mu$ above the unstable
fixed point in the middle (dashed blue lines).}
\label{Pmu}
\end{figure}

For parameter values $(\eta,D)$ within the bistable region and finite
timescale separation $\varepsilon>0$, the coexisting states of excitable
equilibrium and fast oscillations turn into metastable states of the
full system \eqref{eq1a}\textendash \eqref{eq1b}. Based on our slow-fast
analysis, the corresponding dynamics can be understood as follows.
The noisy fluctuations of $\varphi(t)$ around its average distribution,
given by the stationary Fokker-Planck equation \eqref{eq:FoPla},
induces fluctuations of $\langle\sin{\varphi(t)}\rangle_{t}$, and
hence also of $\mu$, around their stationary average values calculated
above. For small $\varepsilon,$ the corresponding distribution of
$\mu$ is centered in narrow peaks at the stable stationary values.
However, with increasing $\varepsilon,$ the nonlinear filtering induces
a strong skewness of each peak in the distribution, and their overlapping
indicates the possibility of noise- induced transitions between the
two metastable states. Figure~\ref{Pmu} shows the distribution for
$\varepsilon=0.005$ and different values of the $\eta$ within the
bistability region. These transitions can be understood in analogy
to the Eyring-Kramers process in a double well potential. In the generic
case of different energy levels for the two potential wells, transitions
in one of the directions occur at a higher rate and the system stays
preferably in state associated to the global minimum of the potential.
Such a behavior of biased switching is very pronounced closed to the
boundaries of the bistability region, where a switching to the state
close to the fold has a much lower probability than switching back.

\begin{figure}[t]
\centering \includegraphics[width=0.99\linewidth]{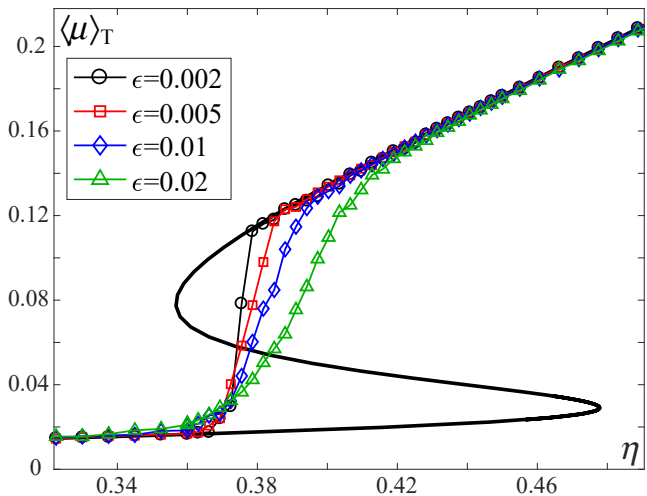} \caption{Long-time averages $\langle\mu\rangle_{T}$ from numerical simulations
of \eqref{eq1a}, \eqref{eq1b} with fixed noise intensity $D=0.008$
and varying control gain $\eta$ at different values of $\varepsilon\in\{0.002,0.005,0.01,0.02\}$.
The black curve represents the corresponding result for the infinite
timescale separation, cf. Fig. \ref{Fig:Cusp}(a).}
\label{fig12}
\end{figure}

\begin{figure}[t]
\centering \includegraphics[width=0.99\linewidth]{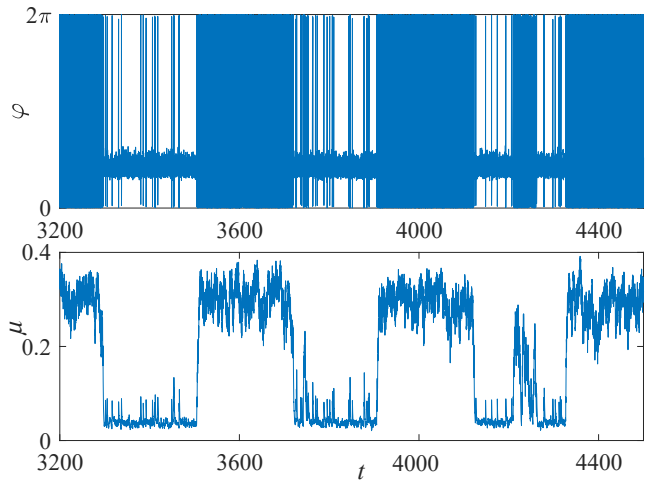} \caption{Time series $\varphi(t)$ (top panel) and $\mu(t)$ (bottom panel)
illustrating the regime of balanced switching. The system parameters
are $\eta=0.38,D=0.008,I_{0}=0.95,\varepsilon=0.01$.}
\label{fig13}
\end{figure}

In Fig.~\ref{fig12} are shown the numerical time averages $\langle\mu(T)\rangle$
for varying control gain $\eta$. One can see that for most values
of $\eta$, the long time behavior is dominated by one of the two
metastable states, which indicates a biased switching process. Nevertheless,
at an intermediate value of $\eta,$ we find a balanced switching,
where transitions in both directions occur at an almost equal rate.
A corresponding time trace is shown in Fig.~\ref{fig13} and Fig.~\ref{fig:1}(b).
For $\varepsilon\to0,$ the switching rate decreases to zero exponentially
and the switching bias in the unbalanced regime increases. This leads
to the characteristic steplike behavior of the averages observed in
Fig.~\ref{fig12} for smaller $\varepsilon$.

The noise-induced switching shown in Fig.~\ref{fig13} and Fig.~\ref{fig:1}(b)
resembles the regime of bursting in neuronal systems. Here it emerges
by an interplay of slow adaptation and noise. In the present setup,
the bursts are triggered just by the stochastic fluctuations. However,
in the regime $\eta>\eta_{\textrm{cu}},$ the system is also quite
susceptible to external inputs, which could initiate the bursts even
without any intrinsic noise.

\section{Discussion and outlook}

\begin{figure}[t]
\centering \includegraphics[width=0.99\linewidth]{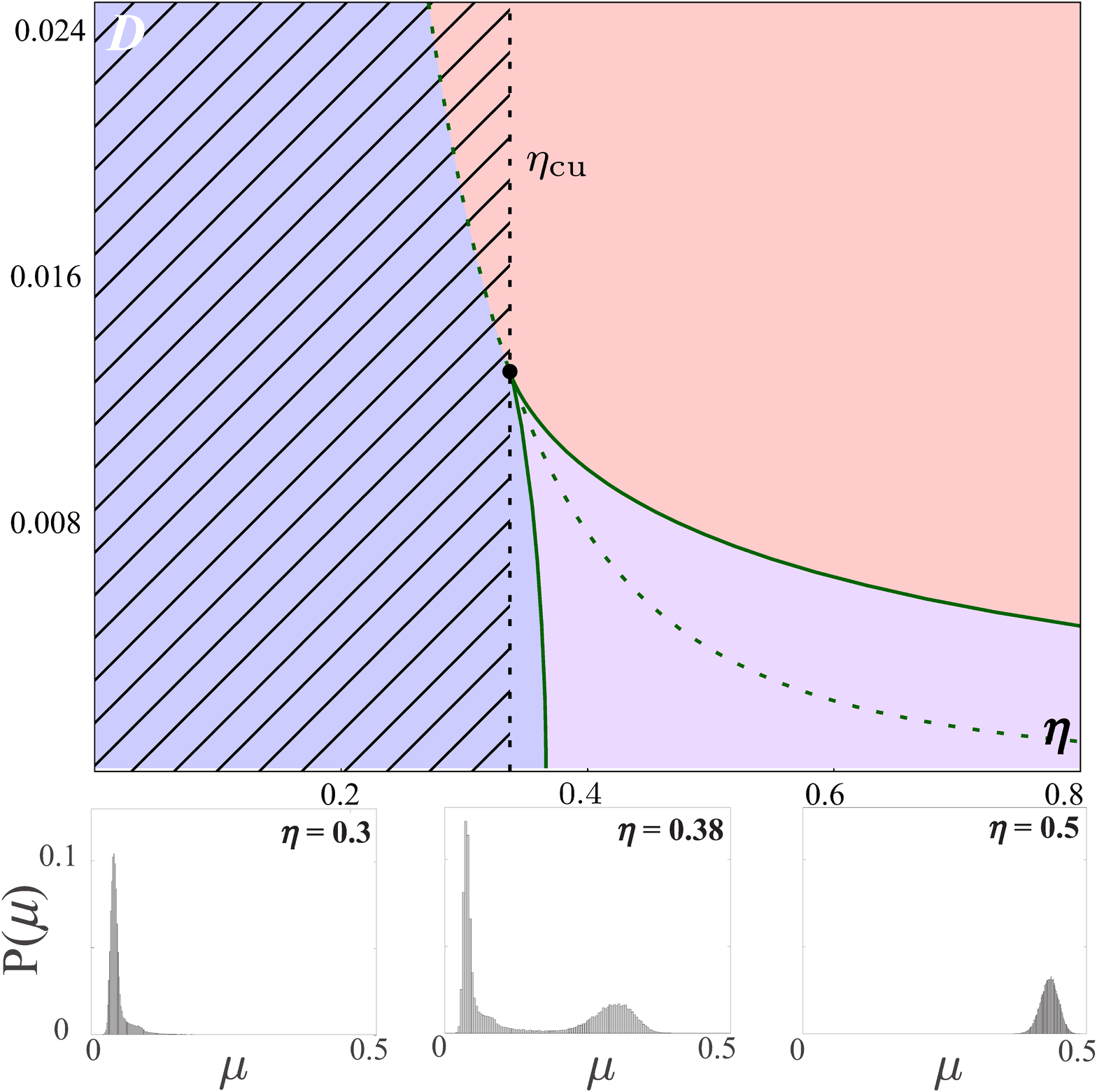} \caption{Upper panel: Parameter regions for different dynamical regimes: noise
induced spiking (blue), noise perturbed oscillations (red), and noise
induced bursting (violet). Enhanced coherence resonance can be found
in the hatched region. Lower panels: Sampled distributions of $\mu(T)$
from numerical solutions with $\varepsilon=0.005,\,D=0.008$ and $\eta\in{0.3,0.38,0.5}$.}
\label{fig14}
\end{figure}
Our model provides a novel perspective on how the dynamics of an excitable
system is influenced by the interaction of a slowly adapting feedback
and noise. The feedback is taken from a low pass filter of a function
that gives a positive feedback to the oscillations by pushing the
excitability parameter towards the oscillatory regime. Since excitability,
feedback, and noise are typical ingredients of neural systems, we
believe that the application of our results to a specific neural model
would be a next natural step, aiming to gain a deeper understanding
of the onset of different dynamical regimes, as well as the means
of controlling their properties and the emerging resonant effects.
In Figure \ref{fig14} are summarized our main results. In particular,
the multiple timescale analysis for the limit of infinite timescale
separation has allowed us to perform a numerical bifurcation analysis
providing the parameter regions for the different dynamical regimes
illustrated in Figure \ref{fig:1}. Numerical simulations for finite
values of $\varepsilon$ (lower panels in Fig. \ref{fig14}) show
that the slowly varying control variable $\mu(T)$ is distributed
around the stationary values from the limiting problem $\varepsilon=0$,
see also Figure \ref{Pmu}. Moreover, we have demonstrated that the
filtered feedback in our model provides an efficient control of the
effect of coherence resonance, which can be substantially enhanced
or suppressed by a corresponding choice of the feedback gain. In the
regime where the limiting problem $\varepsilon=0$ indicates a bistability
between an equilibrium and a fast oscillation, the stochastic fluctuations
at finite values of $\varepsilon$ give rise to a switching between
the associated metastable states. However, our analysis shows that
for sufficiently high noise intensity, this bistability vanishes and
the two different deterministic states can no longer be distinguished.

From the point of view of the theory of multiscale systems, the deterministic
part of the presented model provides one of the simplest examples
combining the regimes of stable equilibrium and oscillations within
the fast subsystem. A rigorous mathematical treatment of the dynamical
transitions between the two regimes and the corresponding reductions
by the standard adiabatic elimination and the averaging technique
is still missing. Also, our approach to analysis of stochastic dynamics
in multiscale systems by introducing a stationary Fokker-Planck equation
for the fast dynamics leads to important questions concerning the
limiting properties of the trajectories and the specific implications
of the fluctuations. Nevertheless, we have considered only the case
when the noise acts in the fast variable. An open problem is to study
how the obtained results are influenced by the noise in the slow variable,
where interesting new effects can be expected \citep{Dannenberg2014}.

\section*{ACKNOWLEDGMENTS\label{ack}}

The work of IF and IB was supported by the Ministry of Education,
Science and Technological Development of the Republic of Serbia under
project No. 171017. SY acknowledges the support from Deutsche Forschungsgemeinschaft
(DFG) under project No. 411803875. The work of MW and SE was supported
by the Deutsche Forschungsgemeinschaft (DFG, German Research Foundation)
- Projektnummer 163436311 - SFB 910.

\section*{Appendix A: Multiscale averaging in the regime of fast oscillations\label{sec:Appendix-A}}

In this appendix we provide a rigorous formal derivation of the slow
averaged equation (\ref{eq:slow2}) for the case of periodic dynamics
in the fast layers.

We apply the following general multiscale Ansatz

\[
\varphi=\bar{\varphi}(t,\varepsilon t)+\varepsilon\hat{\varphi}(t,\varepsilon t),
\]
\[
\mu=\bar{\mu}(t,\varepsilon t)+\varepsilon\hat{\mu}(t,\varepsilon t).
\]
Substituting this Ansatz into \eqref{eq1a}\textendash \eqref{eq1b},
one obtains up to the terms of the order $\varepsilon$
\begin{align*}
\partial_{1}\bar{\varphi}+\varepsilon\partial_{2}\bar{\varphi}+\varepsilon\partial_{1}\hat{\varphi} & =I_{0}-\sin\left(\bar{\varphi}+\varepsilon\hat{\varphi}\right)+\bar{\mu}+\varepsilon\hat{\mu},\\
\partial_{1}\bar{\mu}+\varepsilon\partial_{2}\bar{\mu}+\varepsilon\partial_{1}\hat{\mu} & =\varepsilon\left(-\bar{\mu}-\varepsilon\hat{\mu}+\eta\left(1-\sin\left(\bar{\varphi}+\varepsilon\hat{\varphi}\right)\right)\right),
\end{align*}
where the subscripts 1 and 2 refer to partial derivatives with respect
to $t$ and $\varepsilon t$, respectively. Collecting the terms of
order $\mathcal{O}(1)$, one finds
\begin{equation}
\partial_{1}\bar{\varphi}=I_{0}-\sin\bar{\varphi}+\bar{\mu},\label{eq:phi0}
\end{equation}
\begin{equation}
\partial_{1}\bar{\mu}=0.\label{eq:mubar-1}
\end{equation}
The equation (\ref{eq:mubar-1}) implies that $\bar{\mu}=\bar{\mu}(\varepsilon t)$
depends only on the slow time and acts as a parameter in (\ref{eq:phi0}).
For $\bar{\mu}>1-I_{0}$, equation (\ref{eq:phi0}) has the oscillating
solution $\bar{\varphi}=\varphi_{\bar{\mu}}(t)$ given by (\ref{eq:po}).
Note that the parameters of this solution can depend on the slow time.

As a next step, we consider the terms of order $\varepsilon$:
\begin{align}
\partial_{2}\bar{\varphi}+\partial_{1}\hat{\varphi} & =-\hat{\varphi}\cos\bar{\varphi}+\hat{\mu},\nonumber \\
\partial_{2}\bar{\mu}+\partial_{1}\hat{\mu} & =-\bar{\mu}+\eta\left(1-\sin\bar{\varphi}\right).\label{eq:mubar}
\end{align}
We rewrite Eq.~(\ref{eq:mubar}) as 
\begin{equation}
\partial_{2}\bar{\mu}+\bar{\mu}=-\partial_{1}\hat{\mu}+\eta\left(1-\sin\bar{\varphi}\right),\label{eq:mubar2}
\end{equation}
where the left-hand side depends only on the slow time. Hence, the
solvability condition for (\ref{eq:mubar2}) is the requirement that
its right-hand side is independent on the fast time $t$, i.e. 
\begin{equation}
-\partial_{1}\hat{\mu}+\eta\left(1-\sin\bar{\varphi}\right)=u(T)\label{eq:mutilde}
\end{equation}
with some function $u(T),$where $T=\varepsilon t$ is the slow time.
By integrating (\ref{eq:mutilde}) with respect to the fast time,
we obtain
\begin{equation}
\hat{\mu}(t)=\hat{\mu}(0)+\eta\left(t-\int_{0}^{t}\sin\bar{\varphi}dt\right)-tu(T)\label{eq:tmp}
\end{equation}
The integral in (\ref{eq:tmp}) can be computed using (\ref{eq:phi0}):
\[
\int_{0}^{t}\sin\bar{\varphi}dt=tI_{0}+t\bar{\mu}-\bar{\varphi}(t)+\bar{\varphi}(0)
\]
such that
\[
\hat{\mu}(t)=\hat{\mu}(0)+t\left[\eta\left(1-I_{0}-\bar{\mu}+\frac{\bar{\varphi}(t)-\bar{\varphi}(0)}{t}\right)-u(T)\right]
\]
Taking into account that 
\[
\frac{\bar{\varphi}(t)-\bar{\varphi}(0)}{t}=\Omega(\bar{\mu})+\mathcal{O}\left(\frac{1}{t}\right),
\]
we obtain the expression for $\hat{\mu:}$

\[
\hat{\mu}(t)=\hat{\mu}(0)+t\left[\eta\left(1-I_{0}-\bar{\mu}+\Omega(\bar{\mu})\right)-u(T)\right]+\mathcal{O}(1),
\]
where the linearly growing term must vanish for $\hat{\mu}(t)$ to
be bounded. Setting such a secular term to zero (even without computing
explicitly $\hat{\mu}$), we have 
\[
u(T)=\eta\left(1-I_{0}-\bar{\mu}+\Omega(\bar{\mu})\right),
\]
and, hence, taking into account (\ref{eq:mubar2}) and (\ref{eq:mutilde}),
the equation for the leading order approximation of the slow variable
reads
\[
\partial_{2}\bar{\mu}+\bar{\mu}=\eta\left(1-I_{0}-\bar{\mu}+\Omega(\bar{\mu})\right).
\]
Since $\bar{\mu}$ is the function of the slow time only, we have
$\partial_{2}\bar{\mu}=\bar{\mu}'$ ,which results in the required
averaged equation (\ref{eq:slow2}).

\section*{Appendix B: Explicit solution of the stationary Fokker-Planck equation}

Here we present the analytic solution of the stationary Fokker-Planck
equation (\ref{eq:FoPla})\textendash (\ref{eq:BC}). By integrating
Eq.~(\ref{eq:FoPla}) once one obtains 
\begin{equation}
\frac{D}{2}\partial_{\varphi}\rho-\left(I_{0}+\mu-\sin\varphi\right)\rho=C\label{eq:1ode}
\end{equation}
with a constant $C$ to be determined. Solving (\ref{eq:1ode}), and
taking into account the normalization (\ref{eq:BC}) and the boundary
condition $\rho(0)=\rho(2\pi)$, we arrive at 
\[
\rho(\varphi;\mu,D)=\frac{1}{g_{\Lambda}}\Lambda(\varphi),
\]
where 
\[
\Lambda(\varphi)=\int_{0}^{2\pi}\frac{\Psi(\varphi)}{\Psi(\varphi+\xi)}d\xi,
\]
\[
g_{\Lambda}=\int_{0}^{2\pi}\Lambda(\varphi)d\xi,
\]
\[
\Psi(\varphi)=\exp\left\{ \frac{2}{D}\left[(I_{0}+\mu)\varphi+\cos\varphi-1\right]\right\} .
\]
 \bibliographystyle{apsrev}
\bibliography{refs}

\end{document}